\documentclass{amsart}
\usepackage{amsfonts}

\theoremstyle{definition}

\theoremstyle{remark}

\numberwithin{equation}{section}

\begin{document}

\title{Introduction to Quantum Algorithms}

\author{Peter W. Shor}
\address{AT\&T Labs---Research, Florham Park, NJ 07932, USA }
\email{shor@research.att.com}
\urladdr{http://research.att.com/\~{}shor}

\date{}

\begin{abstract}
These notes discuss the quantum algorithms we know of 
that can solve problems significantly faster than the corresponding
classical algorithms.  So far, we have only discovered a few techniques
which can produce speed up versus classical algorithms.  It is not clear
yet whether the reason for this is that we do not have enough intuition 
to discover more techniques, or that there are only a few problems for 
which quantum computers can significantly speed up the solution.  

In the first section of these notes, I try to explain why the recent
results about quantum computing have been so surprising.  This section
comes from a talk I have been giving for several years now, and discusses 
the history of quantum computing and its relation to the mathematical 
foundations of computer science.  In Sections 
\ref{sec-model}~and~\ref{sec-physics}, I talk about
the quantum computing model and its relationship to physics. 
These sections rely heavily on two of my papers 
[{\sl SIAM J. Comp.\ }{\bf 26} (1997), 1484--1509; 
{\sl Doc.\ Math.\ }{\bf Extra Vol.\ ICM I} (1998), 467--486].
Sections~\ref{sec-simon} and \ref{sec-factoring} illustrate the 
general technique
of using quantum Fourier transforms to find periodicity.
Section~\ref{sec-simon} contains an algorithm of 
Dan Simon showing that quantum computers
are likely to be exponentially faster than classical computers for some
problems.  Section~\ref{sec-factoring} discusses my factoring algorithm, 
which was inspired in part by Dan Simon's paper.  In the final section, 
I discuss Lov Grover's search algorithm, which illustrates a different
technique for speeding up classical algorithms.  These techniques for
constructing faster algorithms for classical problems on quantum computers 
are the only two significant ones which have been discovered 
so far.  
\end{abstract}

\maketitle

%

\section{History and Foundations}
\label{sec-history}

The first results in the mathematical theory of theoretical computer 
science appeared before the discipline of computer science existed; 
in fact, even before electronic computers existed.  
Shortly after G\"odel proved his famous incompleteness result, 
several papers \cite{church,kleene,post,turing} were published
that drew a distinction between computable and non-computable
functions.  These papers showed that there
are some mathematically defined functions which are impossible to
compute algorithmically. Of course, proving such a theorem
requires a mathematical definition of what it means to compute a function.
These papers contained several distinct definitions of computation.  
What was observed was that,
despite the fact that these definitions appear quite different, they
all result in the same class of computable functions.  This led to the
proposal of what is now called the Church-Turing thesis, after two of
its proponents.  This thesis says that
any function that is computable by any means, can
also be computed by a Turing machine.  This is not a mathematical theorem, 
because it does not give a mathematical precise definition of 
computable; it is rather a statement about the real world.  In fact,
many such mathematical theorems have been proven for various definitions
of computation.  What was not widely appreciated until recently is that,
since the Church-Turing thesis implicitly
refers to the physical world, it is in fact 
a statement about physics.  In the sixty years since Church proposed 
his thesis, nobody has discovered any counterexamples to it
and it is now widely accepted.  The current theories of physics  appear to
support this thesis, although as we do not yet have a comprehensive theory 
of physical laws, we must wait until we make a final judgment on this 
thesis.

The model that the majority of these early papers used for intuition 
about computation does not appear to have been a digital computer, as 
these did not yet exist.  Rather, they appear to have been inspired by 
considering a mathematician scribbling on sheets of paper.  
Less than a decade after 1936, the first digital computers 
were built.  As the Church-Turing thesis asserts, the class of functions 
computable by digital machines with arbitrarily large amounts of time and
memory is indeed those functions computable by a Turing machine.  

With the advent of practical digital computers, it became clear that the
distinction between computable and non-computable was much too course for
practical use, as actual computers do not have an arbitrary amount 
of time and memory.
After all, it doesn't do much good in practice
to know that a function is computable
if the sun will burn out long before any conceivable
computer could reach the end of the computation.  What was needed was some
classification of functions as efficiently or inefficiently computable,
based on their computational difficulty.  
In the late 1960's and early 1970's theoretical
computer scientists came up with an asymptotic classification that reflects 
this distinction moderately well in practice, and is also tractable to work 
with theoretically, that is, useful for proving theorems about
the difficulty of computation.  Computer scientists call
an algorithm polynomial-time if the running time grows polynomially in the
input size, and they say that a problem is in the complexity class P if there 
is a polynomial-time algorithm solving it.  
This does not capture the intuitive notion
of efficient perfectly --- hardly anybody would claim that an algorithm 
with an $n^{30}$ running time is feasible --- but it works reasonably 
well in practice.
Experience seems to show that most natural problems in P tend to
have reasonably 
efficient algorithms, and most natural problems not in P tend not to be
solvable much faster than exponential time.   Further, the complexity 
class P has been very useful for proving theorems, an advantage which
is unlikely to hold for any 
definition which differentiates between $O(n^3)$ and $O(n^{30})$ algorithms.  

For the definition of P to make sense, you need to know that it does
not depend on the exact type of computer used for the computation.  This
led to a ``folk'' thesis, 
which we call the polynomial Church's thesis, whose
origins appear to be impossible to pin down, but which has nevertheless
been widely referred to in the literature.  This thesis says that any 
physically computable function can be computed on a Turing machine 
with at most a polynomial increase in the running time.  
That is, if a function 
can be computed on a physical computer in time $T$, it can be computed on a 
Turing machine in time O($T^c$) for some constant $c$ depending only on
the class of computing machine used.

Why might this folk thesis be true?  One explanation might be that the physical
laws of our universe are efficiently simulable by computers.  This would
explain it via the following argument: if we have some physical 
machine that solves a problem, then we can simulate the physical laws 
driving this machine, and by our hypothesis this simulation runs in 
polynomial time.  Conversely, if
we are interested in counterexamples to the polynomial Church's thesis,
we should look at physical systems which appear to be very difficult to
simulate on a digital computer.  Two classes of physical systems immediately
spring to mind
for which simulation currently consumes vast amounts of computer 
time, even while trying to
solve relatively simple problems.  One of these is turbulence, about
which I unfortunately have nothing further to say.  The other is quantum 
mechanics.  

In 1982, Feynman \cite{Feynman}
argued that simulating quantum mechanics inherently required
an exponential amount of overhead, so that it must take enormous amounts
of computer time no matter how clever you are.  This realization was come
to independently, and somewhat earlier, in 1980, in the Soviet Union by 
Yuri Manin~\cite{manin}.
It is not true that all
quantum mechanical systems are difficult to simulate; some of them 
have exact solutions and others
have very clever computational
shortcuts, but it does appear to be true when
simulating a generic quantum mechanics system.  Another thing Feynman
suggested in this paper was the use of quantum com\-pu\-ters to get around 
this.  That
is, a computer based on fundamentally quantum mechanical phenomena 
might be used to simulate quantum mechanics much more efficiently.  
In much the same spirit, you
could think of a wind tunnel as a ``turbulence computer''. 
Benioff~\cite{Benioff} 
had already showed how quantum mechanical processes could be used as the 
basis of a classical Turing machine.  Feynman \cite{Feynman2}
refined these ideas in a later paper.  

In 1985, David Deutsch \cite{Deutsch} gave an abstract
model of quantum computation,
and also raised the question of
whether quantum computers might actually be useful for classical 
problems.  Subsequently, he and a number of other people \cite{DJ,BV,Simon}
came up with
rather contrived-appearing
problems for which quantum computers seemed to work better
than classical computers.  It was by studying these algorithms, 
especially Dan Simon's \cite{Simon}, that I figured out how to 
design the factoring algorithm.

\section{The Quantum Circuit Model}
\label{sec-model}
In this section we discuss the {\sl quantum circuit model} \cite{Yao} 
for quantum computation.  This is a rigorous mathematical model for
a quantum computer.  It is not the only mathematical model that has
been proposed for quantum 
computation; there are also the {\sl quantum Turing machine model} 
\cite{BV,Yao} and the {\sl quantum cellular automata model} \cite{CA1,CA2}.
All these models result in the same class of polynomial-time 
quantum computable functions.  These are, of course, not the only
potential models for quantum computation, and some of the assumptions
made in these models, such as unitarity of all gates, and the lack of
fermion/boson particle statistics, clearly are not physically realistic
in that it is easy to conceive of machines that do not conform to the
above assumptions.  However, there do not seem to be any physically 
realistic models which have more computational power than the ones
listed above.  Neither non-unitarity \cite{AKN} nor fermions
\cite{fermions} add significant power to the mathematical model.
Of these models, the quantum circuit model is possibly the 
simplest to describe. It is also easier to connect with  
possible physical implementations of quantum computers than the quantum 
Turing machine model.   The disadvantage of this model is that it
is not naturally a {\em uniform} model. Uniformity is a technical
condition arising in complexity theory, and to make the
quantum circuit model uniform, additional 
constraints must be imposed on it.  This issue
is discussed later.

In analogy with a classical bit, a two-state quantum system is called
a {\em qubit,} or quantum bit.
Mathematically, a qubit takes a value in the vector space 
${\mathbb C}^2$.  
We single out two orthogonal basis vectors in 
this space, and label these $V_0$ and~$V_1$.  
In Dirac's ``bra-ket'' notation, which comes from physics and
is commonly used in the quantum computing
field, these are represented as $|0\rangle$ and $|1\rangle$.
More precisely, quantum states are invariant under multiplication 
by scalars, so a qubit lives in two-dimensional complex projective 
space.
To conform with physics
usage, we treat qubits as column vectors and operate on them by
left multiplication.

One of the fundamental principles of quantum mechanics is that the joint
quantum state space of two systems is the tensor product of their 
individual quantum state spaces.  Thus, the quantum state space of
$n$ qubits is the space ${\mathbb C}^{2^n}$.  The basis 
vectors of this space are parameterized by binary strings of length $n$.  
We make extensive use of the tensor decomposition of this space 
into $n$ copies of ${\mathbb C}^2$, where we represent a basis state
$V_b$ corresponding to the binary string $b_1b_2\cdots b_n$ by
\[
V_{b_1b_2\cdots b_n} = V_{b_1} \otimes V_{b_2} \otimes
\ldots \otimes V_{b_n}.
\]
In ``bra-ket'' notation, this state is
written as $|b_1b_2b_3\cdots b_n\rangle$ or equivalently, as the
tensor product
$|b_1\rangle |b_2\rangle |b_3\rangle \cdots |b_n\rangle$.
Generally, we use position to distinguish
the $n$ different qubits.  Occasionally we need some other notation
for distinguishing them, in which case we denote the $i$'th qubit
by $V^{[i]}$.  Since quantum states are invariant under multiplication 
by scalars, they can without loss of generality be normalized to be 
unit length vectors; except where otherwise noted, quantum states in 
this paper will be assumed to be normalized.
Quantum computation takes place in the quantum state space of $n$ qubits
${\mathbb C}^{2^n}$,
and obtains extra computational power from its exponential
dimensionality.

In a usable computer, we need some means of giving it the problem we
want solved (input), some means of extracting the answer from it
(output), and some means of manipulating the state of the computer to 
transform the input into the desired output (computation).  We next 
briefly describe input and output for the quantum circuit model.  We 
then take a brief detour to describe the classical circuit model; this will
motivate the rules for performing the computation on a quantum computer.

Since we are comparing quantum computers to classical computers, and
solving classical problems on a quantum computer,
in this paper the input to a quantum computer will 
always be classical information.  
It can thus can be expressed as a binary string $S$ of some length $k$. 
We need to encode this in the initial 
quantum state of the computer, which must be a vector in ${\mathbb C}^{2^n}$.  
The way we do this is to concatenate the bit string $S$ with $n-k$ $0$'s
to obtain the length $n$ string $S0\ldots0$.  We then initialize
the quantum computer in the state $V_{S0\ldots0}$.  Note that the number 
of qubits is in general larger than the input.  These extra qubits,
which we can take to be initialized to $0$, 
are often required for workspace in implementing quantum algorithms. 

At the end of a computation, the quantum computer is in a state
which is a unit vector in ${\mathbb C}^{2^n}$.  This state
can be written explicitly as
$W = \sum_s \alpha_s V_s$ where $s$ ranges over binary strings of 
length $n$, $\alpha_s \in \mathbb{C}$, 
and $\sum_s \left| \alpha_s \right| ^2 = 1$.  These
$\alpha_s$ are called {\em probability amplitudes,} and we say that
$W$ is a {\em superposition} of basis vectors~$V_s$.  In quantum
mechanics, the Heisenberg uncertainty principle tells us that we
cannot measure the complete quantum state of this system.  
There are a large number of permissible measurements; for example,
any orthogonal basis of ${\mathbb C}^{2^n}$ defines a measurement  
whose possible outcomes are the elements of this basis.  However,
we assume that the output is obtained by projecting each qubit 
onto the basis $\{V_0,V_1\}$.  This measurement has the great advantage of
being simple, and it appears that any physically reasonable measurements
can be accomplished by first doing some precomputation and then making
the above canonical measurement.

When applied to a state $\sum_s \alpha_s V_s$, 
this projection produces
the string $s$ with probability $|\alpha_s|^2$.  The quantum
measurement process is inherently probabilistic.  Thus we do not
require that the computation gives the right answer all the time; but
that we obtain the right answer at least $2/3$ of the time. 
Here, the probability $2/3$ can be replaced by any number strictly between
$1/2$ and $1$ without altering the class of functions that
can be computed in polynomial time by
quantum computers---if the probability of obtaining the
right answer is strictly larger than $1/2$, it can be amplified by 
running the computation several times and taking the majority vote 
of the results of these separate computations.

In order to motivate the rules for state manipulation in a quantum circuit, 
we now take a brief detour and describe the classical circuit
model.  Recall that a classical circuit can always be written solely
with the three gates  AND~($\wedge$), OR~($\vee$) and 
NOT~($\neg$).  These three gates are thus said
to form a {\em universal} set of gates.
Besides these three gates, note that we also need elements which 
duplicate the values on wires.  It is arguable that these elements should 
also be classified as gates.  These duplicating ``gates'' are not
possible in the domain of quantum computing, because of the theorem that
an arbitrary quantum state cannot be cloned 
(duplicated) \cite{nocloning1,nocloning2}.

A quantum circuit is similarly built out of logical quantum wires
carrying qubits, and quantum gates acting on these qubits.  Each wire
corresponds to one of the $n$ qubits.  We assume each gate acts on 
either one or two wires.   
The possible physical transformations of a quantum system are unitary 
transformations, so each quantum gate can be described by a unitary matrix.  
A quantum gate on one qubit is then described by
a $2 \times 2$ matrix, and a quantum 
gate on two qubits by a $4 \times 4$ matrix.  Note that 
since unitary matrices are invertible, the computation is
reversible; thus starting with
the output and working backwards one obtains the input.  Further note
that for quantum gates, the dimension of the output space is equal to that
of the input space, so at all times during the computation 
we have $n$ qubits carried on $n$ quantum wires.  

It should be noted that these requirements of unitary and of maintaining only
the original $n$ qubits at all times need to be revised for dealing with
noisy gates, an area not covered in this paper.  In fact, it can be shown that
with these requirements, noisy unitary gates make it impossible to carry out
long computations \cite{ABIN}; some means of eliminating noise
by resetting qubits to values near $0$ is required. 

Quantum gates acting on one or two qubits (${\mathbb C}^2$ or 
${\mathbb C}^4$) naturally induce a transformation on the state space of 
the entire
quantum computer (${\mathbb C}^{2^n}$).  For example, if $A$ is a 
$4 \times 4$ matrix acting on qubits $i$ and $j$, the induced action
on a basis vector of ${\mathbb C}^{2^n}$ is
\begin{equation}
A^{[i,j]}\, V_{b_1b_2\cdots b_n} = \sum_{s=0}^1 \sum_{t=0}^1 \, 
A_{b_i b_j\, s t} \,
V_{b_1 b_2\cdots b_{i-1} s b_{i+1} \cdots b_{j-1} t b_{j+1} \cdots b_n}.
\end{equation}
This is the tensor product of $A$ (acting on qubits $i$ and $j$)
with $n-2$ identity matrices (acting on each of the remaining qubits).
When we multiply a general vector by a quantum gate, it can
have negative and positive coefficients which cancel out, leading to
quantum interference.

As there are for classical circuits, there are universal sets of 
gates for quantum circuits; such a universal set of gates is sufficient
to build circuits for any quantum computation.
One particularly useful universal set of gates is the set of 
all one-bit gates and a specific two-bit gate
called the Controlled {NOT} ({CNOT}).  These gates can efficiently
simulate any quantum circuits whose gates act on
only a constant number of qubits \cite{nine}. 
On basis vectors, the {CNOT} gate negates the second (target)
qubit if and only if the
first (control) qubit is~1.  In other words, it takes 
$V_{XY}$ to $V_{XZ}$ where $Z = X + Y \ (\mathrm{ mod\ }2)$.
This corresponds to the unitary matrix
\begin{equation}
\left(
\begin{array}{rrrr}
1 & 0 & 0 & 0 \\
0 & 1 & 0 & 0 \\
0 & 0 & 0 & 1 \\
0 & 0 & 1 & 0
\end{array}
\right)
\nonumber
\end{equation}

Note that the CNOT is a classical reversible gate.  
To obtain a universal set of classical reversible gates, 
you need at least one reversible
three-bit gate, such as a Toffoli gate; otherwise you can only 
perform linear Boolean computations.  A Toffoli gate is 
a doubly controlled NOT, which negates the 3rd bit if and only if the 
first two are both~1.  By itself the Toffoli gate is universal 
for reversible classical computation, as it can simulate both 
AND and NOT gates \cite{FT}.  Thus, if you can make a Toffoli gate, 
you can perform
any reversible classical computation.  Further, as long as the input
is not erased, any classical computation can be efficiently performed
reversibly \cite{bennett-reverse}, and thus implemented efficiently 
by Toffoli gates.  The matrix corresponding to a Toffoli gate is
\begin{equation}
\left(
\begin{array}{rrrrrrrr}
1 & 0 & 0 & 0 & 0 & 0 & 0 & 0 \\
0 & 1 & 0 & 0 & 0 & 0 & 0 & 0 \\
0 & 0 & 1 & 0 & 0 & 0 & 0 & 0 \\
0 & 0 & 0 & 1 & 0 & 0 & 0 & 0 \\
0 & 0 & 0 & 0 & 1 & 0 & 0 & 0 \\
0 & 0 & 0 & 0 & 0 & 1 & 0 & 0 \\
0 & 0 & 0 & 0 & 0 & 0 & 0 & 1 \\
0 & 0 & 0 & 0 & 0 & 0 & 1 & 0
\end{array}
\right)
\label{toffoli}
\end{equation}

We now define the complexity class BQP, which stands for bounded-error 
quantum polynomial time.  
This is the class of languages which can be computed on a quantum computer
in polynomial time, with the computer giving the correct answer at least
$2/3$ of the time. 

To give a rigorous definition of this complexity
class using quantum circuits, we need to impose uniformity conditions.  
Any specific quantum circuit can only compute a function whose domain
(input) is binary strings of a specific length.
To use the quantum circuit model to implement
functions taking arbitrary length binary strings for input, we need a 
family of quantum circuits, that contains one circuit for inputs of each
length.  Without any further conditions on this family of circuits, 
the designer of this circuit
family could hide an uncomputable function in the 
design of the circuits for each input length.  This definition would
thus result in the unfortunate inclusion of uncomputable functions
in the complexity class BQP{}.  One should note that there is a name for
this nonuniform class of functions.  It is called BQP/poly, meaning that there can be at most a polynomial amount of extra information included in the
circuit design.

To exclude this possibility of including non-computable information in the
circuit,
we require {\em uniformity} conditions on the circuit family.
The easiest way of doing this is to require a classical Turing machine
that on input $n$ outputs a description of the circuit for length 
$n$ inputs, and which runs in time polynomial in~$n$.  
For quantum computing, we need an additional uniformity condition on 
the circuits.  It is also be possible for the circuit designer 
to hide uncomputable (or hard-to-compute) information in the unitary 
matrices corresponding to quantum gates.  We thus require that the
$k$'th digit of the entries of these matrices can be computed by a
second Turing machine in
time polynomial in~$k$ and $n$.  Although we do not have space to discuss this
fully, the power of the classical machines designing the circuit family
can actually be varied over a wide range; they can be varied from classes
much smaller than P to the classical randomized class BPP.  
This helps us convince 
ourselves that we have the right definition of BQP{}.

The definition of polynomial time computable functions on a quantum 
computer is thus those functions computable
by a {\em uniform} family of circuits whose size (number of gates)
is polynomial in the length of the input, and which for any input gives 
the right answer at least $2/3$ of the time.  The corresponding set of 
languages (languages are functions with values in $\{0,1\}$) is called BQP{}.

\section{Relation of the Model to Quantum Physics}
\label{sec-physics}
The quantum circuit model of the previous section
is much simplified from the realities of
quantum physics.  There are operations possible in physical
quantum systems
which do not correspond to any simple operation allowable in the
quantum circuit model, and complexities that occur when 
performing experiments that are not reflected 
in the quantum circuit model.  This section contains a brief 
discussion of these issues, some of which are discussed more 
thoroughly in \cite{BV,recentDiVincenzo}.

In everyday life, objects behave very classically, and on large
scales we do not see any quantum mechanical behavior.  This is
due to a phenomenon called decoherence, which makes
superpositions of states decay, and makes large-scale superpositions
of states decay very quickly.
A thorough, elementary, discussion of decoherence can be found in \cite{Zurek};
one reason it occurs is that we are dealing with open systems rather
than closed ones.  Although closed
systems quantum mechanically undergo unitary evolution,
open systems need not.  They are subsystems of systems undergoing 
unitary evolution, 
and the process of taking subsystems does not preserve unitarity.

However hard we may try to isolate quantum computers from
the environment, it is virtually inevitable
that they will still undergo some decoherence and
errors.  We need to know that these processes do not fundamentally 
change their behavior.  Using no error correction, if each gate results 
in an amount of decoherence and error of order $1/t$, then
$O(t)$ operations can be performed
before the quantum state becomes so noisy as to usually give the 
wrong answer~\cite{BV}.  
Active error correction can improve this situation substantially;
this is discussed in Gottesman's notes for this course \cite{Gottesman3}.

In some proposed physical architectures for quantum computers,
there are restrictions that are more severe than the quantum circuit
model given in the preceding section.  
Many of these restrictions do not change the class 
BQP{}.  For example, it might be the case that a gate could 
only be applied to a pair of {\em adjacent} qubits.  
We can still operate on a 
pair of arbitrary qubits: by repeatedly exchanging one of these qubits 
with a neighbor we can bring this pair together.  If there are $n$ qubits
in the computer, this can only increase the computation time by a factor
of $n$, preserving the complexity class BQP{}.

The quantum circuit model described in the previous section postpones
all measurements to the end, and assumes that we are not allowed
to use probabilistic steps.  Both of these possibilities are allowed
in general by quantum mechanics, but neither possibility makes the
complexity class BQP larger \cite{BV}.  For fault-tolerant quantum
computing, however, it is very useful to permit measurements in the 
middle of the computation, in order to measure and correct errors.

The quantum circuit model also assumes that we only operate on a
constant number of qubits at a time.  In general quantum systems,
all the qubits evolve simultaneously according to some
Hamiltonian describing the system.  This simultaneous evolution of
many qubits cannot be described by a single gate in our model,
which only operates on two qubits at once.
In a realistic model of quantum computation, however, we cannot allow
general Hamiltonians, since they are not experimentally
realizable.  Some Hamiltonians that act on all the qubits at once
are experimentally realizable.  
It would be nice to know that even though these
Hamiltonians cannot be directly 
described by our model, they cannot be used to 
compute functions not in BQP in polynomial time.  
This could be accomplished by showing that systems with such 
Hamiltonians can be efficiently simulated by a quantum computer.  
Some work has been done on simulating Hamiltonians on quantum
computers \cite{AL,Lloyd,Zalka}, but I do not believe this question
has been completely addressed yet.

\section{Simon's Algorithm}
\label{sec-simon}
In this section, we give Dan Simon's algorithm \cite{Simon} for a problem 
that takes exponential time on a classical computer, but quadratic time
on a quantum computer.  This is an ``oracle'' problem, in that there
is a function $f$ given as a ``black box'' subroutine, and the computer is
allowed to compute $f$, but is not allowed to look at the code for $f$.
In fact, to prove the lower bound on a classical computer, we
must permit the computer to use functions $f$ which are not efficiently  
computable.  

We now describe Simon's problem.  The computer is given a function 
$f$ mapping ${\bf F}_2^n$ to ${\bf F}_2^n$ which has the property that
there is a $c$ such that
\begin{equation}
f(x) = f(y) \longleftrightarrow x \equiv y + c \ \ \ (\mathrm{mod\ }{\bf F}_2^n)
\label{F2period}
\end{equation}
Here, the addition is bitwise binary addition.  
Essentially, this is a function which is periodic over ${\bf F}_2^n$ with
period $c$.

We now describe the lower bound for a classical computer.  Suppose that 
the function $f$ is chosen at random from all functions with property
(\ref{F2period}).   We show that you need to compute $O(2^{n/2})$ function
evaluations to find $c$.
Suppose that you have evaluated $s$ values of $f$.  You have then eliminated
at most one value of $c$ for each pair of the $s$ values of $f$ computed,
but $c$ is equally likely to be any of the remaining possibilities.
Thus, after computing $s$ values of $f$, you will have eliminated at most
$s(s-1)/2$ values of $c$.  At least half the time, 
you must try more than half the possibilities for $c$, and this takes 
$O(2^{n/2})$ function evaluations.

We now describe Simon's algorithm for finding the period on a quantum 
computer.  To do this, we need to introduce the Hadamard gate,
\[
H = \frac{1}{\sqrt{2}}\left(
\begin{array}{cc}1&1\\1&-1\end{array}\right).
\]
Now, suppose that we apply the Hadamard transformation to each of $k$
qubits.  We obtain, for a vector $a$ in ${\bf F}_2^k$, 
\begin{equation}
H^{\otimes k}(V_a) = \frac{1}{2^k/2} \sum_{b=0}^{2^k-1} (-1)^{a\cdot b} V_b.
\end{equation}
It is easy to see that each entry of the matrix $H^{\otimes k}$ is
$\pm 2^{-k/2}$.  Further, the $(a,b)$ entry picks up a factor of $-1$ 
for each position which is 1 in both $a$ and $b$, giving a sign of 
$(-1)^{a\cdot b}$.  Here,
\[
a\cdot b = \sum_i a_i b_i \ \ \ (\mathrm{mod\ }2)
\]
is the binary inner product of $a$ and $b$.
This is in fact the Fourier transform over ${\bf F}_2^k$.

We are now ready to describe Simon's algorithm.  We will use two registers,
both with $n$ qubits.  We start with the state $V_0 \otimes V_0$.  The first
step is to take each qubit in the first register
to $\frac{1}{\sqrt{2}}(V_0+V_1)$, putting the first register in an equal 
superposition o all binary strings of length $n$.  The 
computer is now in the state
\[
2^{-n/2} \sum_{x=0}^{2^n-1} V_x \otimes V_0.
\]
The second step is to compute $f(x)$ in the second register.  We now
obtain the state
\[
2^{-n/2} \sum_{x=0}^{2^n-1} V_x \otimes V_{f(x)}.
\]
Note that since the input $x$ of the function $f(x)$ is kept in memory,
this is a reversible classical transformation, and thus unitary.  
The third step is to take the Fourier transform of the first register.
This leaves the first register in the state
\[
2^{-n} \sum_{x=0}^{2^n-1} \sum_{y=0}^{2^n-1} (-1)^{x \cdot y} 
V_y \otimes V_{f(x)}.
\]
Finally, we observe the state of the computer in the basis ${V_i}\otimes V_j$.
We see the state $V_y \otimes V_{f(x)}$ with probability equal to the square
of its amplitude in the above sum.  
There are exactly two $x$ which give the value $f(x)$, namely $x$ and $x+c$.
The probability of observing $V_y \otimes V_{f(x)}$ is thus
\[
2^{-2n} \left( (-1)^{x\cdot y} + (-1)^{(x+c)\cdot y} \right)^2. 
\]
This probability is either $2^{2n - 2}$ or $0$, depending on whether
$y \cdot c$ is 0 or~1.  The above measurement thus produces a random $y$ 
with $c \cdot y = 0$.  It is straightforward to show that $O(n)$ such $y$'s
chosen at random will be of full rank in $c^\perp$, the $n-1$ dimensional
space perpendicular to $c$, and thus determine $c$ uniquely.  Thus, if we
repeat the above procedure $O(n)$ times, we will be able to deduce $c$.
Since each of these repetitions takes $O(n)$ steps on the quantum computer,
we obtain the answer in $O(n^2 + nF)$ time, where $F$ is the cost of the
evaluating the function $f$.

Simon's algorithm is at least a moderately convincing argument that 
BQP is strictly larger than BPP, although 
it is not a rigorous proof.  However, Simon's problem is contrived in 
that it does not seem to have
arisen in any other context.
It did point the way to my discovery of the factoring algorithm, which
will be discussed in the next section.  The factoring algorithm is a much 
less convincing argument that BQP is larger than BPP, as nobody really 
knows the complexity of factoring.  However, as factoring is a widely
studied problem that is fundamental for public key cryptography \cite{RSA}, 
the quantum factoring algorithm brought widespread attention to the field
of quantum computing.

\section{The Factoring Algorithm}
\label{sec-factoring}
For factoring an $L$-bit number $N$, the best classical algorithm known
is the number field sieve \cite{NFS}; this algorithm asymptotically takes time
$O(\exp (c L^{1/3} \log^{2/3} L) )$.
On a quantum computer, the quantum factoring algorithm takes asymptotically
$O(L^2 \log L \log \log L)$ steps.  The key idea of the quantum
factoring algorithm is the use of 
a Fourier transform to find the period of the sequence
$u_i = x^i \ (\mathrm{ mod\ }N)$, from which period a factorization
of $N$ can be obtained.
The period of this sequence is exponential in $L$, so this approach is not
practical on a digital computer.  On a quantum computer, however, we
can find the period in polynomial time by exploiting the $2^{2L}$-dimensional
state space of $2L$ qubits, and taking a Fourier transform over this
space.  The exponential dimensionality of this space permits us to take 
the Fourier transform of an exponential length sequence.  How this works
will be made clearer by the following sketch of the algorithm, the
full details of which are in \cite{Shor-factoring}, along with a 
quantum algorithm for finding discrete logarithms.

The idea behind all the fast factoring algorithms (classical or
quantum) is fairly simple.  To factor $N$, find two residues mod $N$
such that 
\begin{equation}
s^2 \equiv t^2 \ (\mathrm{ mod\ } N)
\end{equation}
but $s \not\equiv \pm t \ (\mathrm{ mod\ } N)$.  We now have
\begin{equation}
(s+t)(s-t) \equiv 0 \ (\mathrm{ mod\ } N)
\end{equation}
and neither of these two factors is $0 \ (\mathrm{ mod\ } N)$.  
Thus, $s+t$ must contain one
factor of $N$ (and $s-t$ another).  We can extract this 
factor by finding the greatest common divisor of $s+t$ and $N$; this
computation can be done in polynomial time using Euclid's algorithm.

In the quantum factoring algorithm, we find the multiplicative period~$r$
of a residue $x \ (\mathrm{mod\ } N)$.  This period $r$ satisfies
$x^r \equiv 1 \ (\mathrm{mod\ } N)$. If we are lucky and $r$ is even, then
both sides of this congruence are squares and we can try the above
factorization method.  If we are just a little bit
more lucky, then $x^{r/2} \not\equiv -1 \ (\mathrm{mod\ } N)$, and we 
obtain a factor by computing $\mathrm{gcd}(x^{r/2}+1,N)$.  The greatest
common divisor can be computed in polynomial time on
a classical computer using Euclid's algorithm.

It is a relatively
simple exercise in number theory to show that for large $N$
with two or more prime factors, at least half the residues 
$x \ (\mathrm{ mod\ } N)$ produce prime factors using this technique, and 
that for most large $N$ the fraction of good residues $x$ is much 
higher; thus, if we try several different values for $x$, we have to 
be particularly unlucky not to obtain a factorization using this method.

We now need to explain what the quantum Fourier transform is.
The quantum Fourier transform on $k$ qubits
maps the state $V_a$, where $a$ is considered as an integer between
$0$ and $2^k-1$, to a superposition of the states $V_b$ as follows:
\begin{equation}
V_a \rightarrow \frac{1}{2^{k/2}}
\sum_{b=0}^{2^k-1} \exp(2 \pi i a b / 2^k)\, V_b
\label{dft}
\end{equation}
It is easy to check that this transformation defines a unitary matrix.
It is not as straightforward to implement this Fourier
transform as a sequence of one- and two-bit quantum gates.  However,
an adaption of the Cooley-Tukey algorithm decomposes this
transformation into a sequence of $k(k-1)/2$ one- and two-bit gates.  
More generally, the discrete Fourier transform over any 
product $Q$ of small primes (each of size at most $\log Q$) can be 
performed in polynomial time on a quantum computer.  We will show
how to break the above Fourier transform of Eq.~(\ref{dft})
into this product of two-bit
gates at the end of this section.

We are now ready to give the quantum algorithm for factoring.  What
we do is design a polynomial-size circuit which starts in the quantum
state $V_{00 \ldots 0}$ and whose output, with reasonable probability,
lets us factor an $L$-bit number $N$ in polynomial time using a
digital computer.  This circuit has two main registers, the first of 
which is composed of $2L$ qubits and the second of $L$ qubits.  It also
requires a few extra qubits of work space, which we do not
mention in the summary below but which are required for implementing
the step (\ref{exponentiation}) below.

We start
by putting the computer into the state representing the superposition
of all possible values of the first register:
\begin{equation}
\frac{1}{2^L} \sum_{a=0}^{2^{2L}-1} V_a \otimes V_0.
\end{equation}
This can easily be done using $2L$ gates by putting each of the qubits in 
the first register into the state $\frac{1}{\sqrt{2}}(V_0+V_1)$.

We next use the value of $a$ in the first register to compute the value
$x^a \ (\mathrm{ mod\ }N)$ in the second register.  This can be done using
a reversible classical circuit for computing $x^a \ (\mathrm{ mod\ }N)$ 
from~$a$.  Computing $x^a \ (\mathrm{ mod\ }N)$ using repeated squaring 
takes 
$O(L^3)$ quantum gates using the grade school multiplication algorithm,
and asymptotically $O(L^2 \log L \log \log L)$ gates using fast integer 
multiplication
(which is actually faster only for moderately large values of~$L$).  
This leaves the computer in the state
\begin{equation}
\frac{1}{2^L} \sum_{a=0}^{2^{2L}-1} V_a \otimes V_{x^a (\mathrm{ mod\ }N)}.
\label{exponentiation}
\end{equation}

The next step is to take the discrete Fourier transform of the first 
register, as in Equation (\ref{dft}).  This puts the computer into 
the state
\begin{equation}
\frac{1}{2^{2L}} \ \sum_{a=0}^{2^{2L}-1} \  \sum_{c=0}^{2^{2L}-1} 
\exp(2 \pi i ab / 2^{2L}) V_c \otimes V_{x^a (\mathrm{ mod\ }N)}.
\label{end-computation}
\end{equation}

Finally, we measure the state of our machine.  This yields the output 
$V_c \otimes V_{x^j (\mathrm{ mod\ }N)}$ with probability equal
to the square of the coefficient on this state in the 
sum~(\ref{end-computation}).  Since many values of $x^a \ (\mathrm{mod\ }N)$
are equal, many terms in this sum contribute to each coefficient.
All these $a$'s giving the same 
value of $x^a \ (\mathrm{mod\ }N)$ can be represented
as 
\[
a = a_0 + br,
\] 
where $a_0$ is the smallest of these $a$'s and $b$ is some integer
between $0$ and $\lceil 2^{2L}/r \rceil$. 
Explicitly, this probability is:
\begin{equation}
\frac{1}{2^{4L}} \left| 
\exp(2 \pi i  a_0 c   / 2^{2L}) 
\sum_{b=0}^{\lfloor 2^{2L} / r \rfloor + \eta} 
\exp(2 \pi i  b rc  / 2^{2L}) \right|^2.
\label{factor-prob}
\end{equation}
where $\eta$ is either $0$ or $1$, depending on the values of 
$2^{2L}\ (\mathrm{mod\ }r)$ and $a_0$.
This sum in Eq.~(\ref{factor-prob}) is a geometric sum of unit complex 
numbers equally spaced around the
unit circle, and it is straightforward to check that 
this sum is small except when these complex numbers point predominantly in
the same direction.  For this to happen, we need that 
the angle between the two complex phases for $b$ and $b+1$ is on the order
of the reciprocal of the number of possible $b$'s, i.e., that
\begin{equation}
r c / 2^{2L} = d + O(r/ 2^{2L})
\label{easy-eq}
\end{equation}
for some integer $d$.  We thus are likely to observe only values of 
$b$ satisfying (\ref{easy-eq}).  Recalling that $2^{2L} \approx N^2$, we
can rewrite this equation to obtain
\begin{equation}
\frac{c}{2^{2L}} = \frac{d}{r} + O(1/N^2).
\label{approx-eq}
\end{equation}
We know $c$ and $2^{2L}$, and we want to find $r$.
Since both $d$ and $r$ are less than $N$, if the $O(1/N^2)$ in 
Eq.~(\ref{approx-eq}) were exactly $1/2N^2$, we would have
\[
\left| \frac{c}{2^{2L}} - \frac{d}{r}\right| \leq \frac{1}{2N^2}
\]
and $\frac{d}{r}$ would be the
closest fraction to $c/2^{2L}$ with numerator and denominator
less than $N$.  In actuality, it is likely to be one of the
closest ones.  Thus, all we need do to find $r$ is to round $c/2^{2L}$ 
to find all close fractions with denominators less than $N$.  
This can be done in 
polynomial time using a continued fraction expansion, and since we can
check whether we have obtained the right value of $r$, we can search the
close fractions until we have obtained the correct one.
We chose $2L$ as the size of the first register in order to make 
$d/r$ likely to be the closest fraction to $c/2^{2L}$ with numerator
and denominator at most $N$. 

More details of this algorithm can be found in \cite{Shor-factoring}. 
Recently, Zalka \cite{Zalka-factoring} has analyzed the resources 
required by this algorithm much more thoroughly, improving upon their
original values in many respects.  For example, he shows that you can
use only $3L+o(L)$ qubits, whereas the original algorithm required
$2L$ extra qubits for workspace, giving a total of $5L$ qubits.  He
also shows how to efficiently parallelize the algorithm to run on a
parallel quantum computer.

\subsection{Implementing the Quantum Fourier Transform}
We now show how to break the discrete Fourier transform
(Eq.~\ref{dft}) into a product of two-bit gates, a step which we previously
postponed to this subsection.  Let us consider the Fourier
transform on $k+1$ bits.
\begin{equation}
V_a \rightarrow \frac{1}{2^{(k+1)/2}}
\sum_{b=0}^{2^{k+1}-1} \exp(2 \pi i a b / 2^{k+1})\, V_b
\end{equation}
We will assume that we have an expression for the Fourier transform
on $k$ qubits, and show how to obtain an expression for the Fourier 
transform on $k+1$ qubits using only $k+1$ additional gates.
 
We break the input space $V_a$ on $k+1$ qubits into the tensor 
product of a $k$-qubit space and a $1$-qubit space, so that 
$V_a = V_{a_-} \otimes V_{a_0}$, where the $(k+1)$-bit string $a$ is 
the concatenation of 
the $k$-bit string $a_-$ and
the one-bit string $a_0$. 
Thus, $a_0$ is the rightmost bit of the binary number~$a$, i.e., 
the units bit.  
We similarly break the output space $V_b$ into the tensor product of a 
$1$-qubit space and
a $k$-qubit space, but this time we choose the first bit 
as the $1$-qubit space,
so $V_b = V_{b_k} \otimes V_{b_-}$, where $b_k$ is the leftmost bit of $b$,
i.e. the bit with value $2^k$, and $b_-$ comprises the $k$ rightmost bits.
Now, the Fourier transform becomes
\begin{equation}
V_{a_-} V_{a_0} \rightarrow 2^{\frac{k+1}{2}}
\sum_{{a_k = 0 \atop b_0 = 0}}^1 \sum_{a_- = 0 \atop b_- =0}^{2^{k}-1} 
\exp\left(2 \pi i 
{\textstyle
\big(\frac{a_0 b_k}{2} + \frac{a_0 b_-}{2^{k+1}} 
+ {\scriptstyle a_- b_k} + \frac{a_- b_-}{2^k}\big)}\right)
\, V_{b_k} V_{b_-} .
\end{equation}
We now analyze this expression.  First, the term
$\exp(2 \pi i a_- b_k )$ is always~$1$, and thus can be dropped.  The
term $\exp(2 \pi i a_- b_- / 2^{k})$ is the phase factor in the 
quantum Fourier transform on $k$ qubits.  
Thus, if we first perform the Fourier transform on $k$ qubits (which we
can do by the induction hypothesis), 
we take $V_{a_-}$ to $V_{b_-}$ and 
obtain this phase factor.
The term $\exp(2 \pi i a_0 b_- / 2^{k+1})$ can be expressed as the product
of $k$ gates, by letting the gate
\[
T_{j,k} = 
\left(
\begin{array}{cccc}
\ 1\ & 0 &0&0 \\
0 & \ 1\ &0&0 \\
0 & 0 &\ 1\ &0 \\
0&0&0 & \exp\left(\frac{2\pi i}{ 2^{k+1-j}} \right)
\end{array}
\right)
\]
operate on the qubits corresponding to $a_0$ and $b_j$, by which
we mean the bit of $b_-$ with value $2^j$, i.e., the $j+1$'st bit from the
right.
This gate applies the phase factor of 
$\exp(2\pi i/2^{k+1-j})$ if and only if both the bits $a_0$ and $b_j$ are 1.
Finally, the term 
\[
\exp(2 \pi i a_0 b_k / 2) = (-1)^{a_0 \cdot b_k}
\]
is the unitary gate
\[
H = \frac{1}{\sqrt{2}}
\left(
\begin{array}{cc}
1 & 1 \\
1 & -1 
\end{array}
\right)
\]
which
takes $V_{a_0}$ to $V_{b_k}$ with the phase factor $(-1)^{a_0 \cdot b_k}$.
We now see that we can obtain the Fourier transform on $k+1$ qubits by
first applying the Fourier transform on $k$ qubits, taking $V_{a_-}$ to
$\sum \exp(2\pi i a_-b_-/2^k) V_{b_-}$, next applying the
gate $T_{j,k}$ on the qubits $V_{a_0}$ and $V_{b_j}$ for $j = 0$ to $k-1$, and 
finally by applying the gate $H$ on the qubit $V_{a_0}$ (yielding in the
qubit $b_k$.  For those readers who are familiar
with the Cooley-Tukey fast Fourier
transform, this is almost a direct translation of it to a quantum algorithm. 
Multiplying the gates $T_{j,k}$ for a fixed $k$ gives the ``twiddle factor''
of the Cooley-Tukey FFT.  

One objection that might be raised to this expansion of the Fourier transform 
is that it requires gates with exponentially small
phases, which could not possibly be implemented with any physical accuracy.
In fact, one can omit these gates and obtain an approximate Fourier transform
which is close enough
to the actual Fourier transform that it barely changes the
probability that the 
factoring algorithm succeeds \cite{Coppersmith}.  
This reduces the number of gates
required for the quantum Fourier transform from $O(k^2)$ to $O(k \log k)$.  

\section{Grover's Algorithm}
\label{sec-grover}

Another very important algorithm in quantum computing is L.~K.
Grover's search algorithm, which searches an unordered
list of $N$ items (or the range of an efficiently computable function)
for a specific item in time $O(\sqrt{N})$, an improvement on the 
optimal classical algorithm, which must look
at $N/2$ items on average before finding a specific item \cite{Grover1}.  
The technique used in this algorithm can be applied to a number of other
problems to also obtain a square root speed-up \cite{Grover2}.  If you
are searching an unordered database, this square root speed-up is as
good as a quantum computer can do; this is proved using techniques
developed in \cite{BBBV}.  Finally, a generalization of both Grover's 
search algorithm and the lower bound above gives tight bounds on how much
a quantum computer can amplify a quantum procedure that has
a given probability of success \cite{BCDZ}.  The quantum search algorithm
can be thought of in these terms; the procedure is just that of choosing
a random element of the $N$-element list, so the probability of success 
is $1/N$.  A quantum computer can amplify this probability to near-unity
by using $O(\sqrt{N})$ iterations while a classical computer requires order
$N$ iterations.  I sketch Grover's search algorithm below.

Grover's algorithm uses only three transformations.  The first is
the transformation $W = H^{\otimes k}$, which is the transformation
obtained by applying the matrix
\[
H = \frac{1}{\sqrt{2}} \left( \begin{array}{cc}
1&1\\
1&-1
\end{array}
\right)
\]
to each qubit.  It is easy to check that $W^2 = \mathrm{Id}$, 
because $H^2 = \mathrm{Id}$.  The second transformation is $Z_0$, which 
takes the basis vector $V_0$ to $-V_0$ and leaves $V_i$ unchanged for 
$i \neq 0$.  The third is $Z_t$, which takes $V_t$ to $-V_t$ and leaves 
$V_i$ unchanged for $i \neq t$, where the $t$'th element of the list is 
the one we are trying to find.  At first glance, it might seem that we need 
to know $t$ to apply $Z_t$; however, if we can design a quantum circuit
that tests whether an integer $i$ 
is equal to $t$, than we can use it to perform the transformation~$Z_t$.  
For example, if we are searching for a specific element in an 
unordered list, it is fairly straightforward
to write a program that tests whether the 
$i$'th element of the list is indeed the desired element, and negates the
phase if it is, without knowing where the desired
element is in the list.  Similarly, if we are searching for a solution to 
some mathematical problem, we need only to be able to 
efficiently test whether a given 
integer $i$ encodes a solution to the problem. 

Suppose that we are searching among $N=2^k$ items, which are encoded
by the integers $0$ to $N-1$.  Here we use $k$ qubits to keep track
of the items.  We will now calculate that if we start in the superposition
\[
\Sigma_{i=0}^{N-1} \alpha_i V_i
\]
then the transformation $- WZ_0W$ leaves us in the state
\[
\Sigma_{i=0}^{N-1} (2m-\alpha_i) V_i
\]
where $m = \frac{1}{N} \Sigma_0^{N-1} \alpha_i$ is the mean of all the 
amplitudes.  The proof of this follows from the observation that after the
transform $W$, the amplitude of $V_0$ is $\sqrt{N} m$.  
Recall that $W^2 = \mathrm{Id}$.  These two observations can be used to
show that the transformation $W Z_0 W$ extracts the mean $m$ in the amplitude
of $V_0$, negates it, and redistributes it negated over all the
basis states $V_i$.  The transformation $WZ_0W$ thus 
takes $\Sigma_i \alpha_i V_i$ to $\Sigma_i (\alpha_i - 2m) V_i$. 

We are now in a position to describe Grover's algorithm in detail.
We start in the equal superposition of all $V_i$, i.e. the state
$\frac{1}{\sqrt{N}}\Sigma_{i=0}^{N-1} V_i$.
We then repeat the transformation $Z_t W Z_0 W$ for $c \sqrt{N}$ iterations,
for the appropriately chosen constant~$c$.  What this accomplishes is to
gradually increase the amplitude on $V_t$ at the expense of all the other 
amplitudes, until after $c \sqrt{N}$ iterations
the amplitude on $V_t$ is nearly unity.  Suppose
that we have reached a point where the amplitude on $V_i$ is $\alpha$ for
all $i \neq t$ and $\beta$ for $V_t$.  It is easy to see that in the
next step, these amplitudes are $2m-\alpha$ and $\beta + 2m$, respectively,
where $m = (\beta + (N-1) \alpha)/N$ is the mean amplitude.
When $\beta$ is small, $m \approx \alpha \approx 1/\sqrt{N}$, and thus 
the amplitudes on $V_i$, $i \neq t$ decrease slightly and the amplitude 
on $V_t$ increases by approximately $2/\sqrt{N}$.  I will not go
into the details in this write-up, but this at least gives the intuition that,
after $c \sqrt{N}$ steps, we obtain a state very close to $V_t$.  
There are many variations of this algorithm, including ones that work 
when there is more than one desired solution. 
For more details, I recommend reading Grover's paper \cite{Grover1}.

Finally, as Feynman suggested, it appears that quantum computing is good
at computing simulations of quantum mechanical dynamics.  I will not
be discussing this.  Some work in this regard has appeared in 
\cite{AL,Lloyd,Zalka}, but much remains to be done.

\end{document}